 \documentclass[final, 5p,times,twocolumn]{elsarticle2}
 
\biboptions{sort&compress}

\usepackage{amssymb}

\usepackage{textcomp}
\usepackage{booktabs, multirow}
\usepackage[colorlinks, citecolor=blue, linkcolor=blue,]{hyperref}

\usepackage[labelfont=bf, labelsep=period]{caption}
\usepackage{flushend}

\journal{Current Opinion in Colloid \& Interface Science}

\begin{document}

\begin{frontmatter}


\title{Capillary suspensions: Particle networks formed through the capillary force}
\author{Erin Koos}
\ead{erin.koos@kit.edu}
\address{Institute for Mechanical Process Engineering and Mechanics, Karlsruhe Institute of Technology (KIT), Stra{\ss}e am Forum 8, 76131 Karlsruhe, Germany}

\begin{abstract}
The addition of small amounts of a secondary fluid to a suspension can, through the attractive capillary force, lead to particle bridging and network formation. The capillary bridging phenomenon can be used to stabilize particle suspensions and precisely tune their rheological properties.  This effect can even occur when the secondary fluid wets the particles less well than the bulk fluid. These materials, so-called capillary suspensions, have been the subject of recent research studying the mechanism for network formation, the properties of these suspensions, and how the material properties can be modified. Recent work in colloidal clusters is summarized and the relationship to capillary suspensions is discussed. Capillary suspensions can also be used as a pathway for new material design and some of these applications are highlighted. Results obtained to date are summarized and central questions that remain to be answered are proposed in this review. 
\end{abstract}

\begin{keyword}
Capillary suspensions \sep Ternary systems \sep Capillary force \sep Percolation \sep Colloidal clusters

\emph{Article history:}~ Received 31 July 2014; Received in revised form 20 October 2014; Accepted 21 October 2014



\end{keyword}

\end{frontmatter}



\section{Introduction} \label{sec:intro}
Ternary particle-liquid-liquid systems composed of particles dispersed in two immiscible liquids can form a variety of structures depending on the ratio of the three components and their material properties.  The particles can stabilize emulsions forming Pickering emulsions~\cite{Aveyard:2003}, which cluster together forming spherical agglomerates that readily separate from a bulk fluid~\cite{Cattermole:1904, Leonard:1991}, or prevent the spinodal decomposition of the two fluids forming a Bijel~\cite{Stratford:2005}. Recently, it was determined that when a small amount of the second immiscible liquid is added to the continuous phase of a suspension, the rheological properties of the mixture are dramatically altered from a fluid-like to a gel-like state or from a weak to a strong gel~\cite{Koos:2011}.  This transition increases the yield stress and viscosity by several orders of magnitude as the volume fraction of the second fluid increases and is attributed to the capillary forces of the two fluids on the solid particles.  Thus, these systems have been termed \emph{capillary suspensions}.  Capillary suspensions are a new class of materials that can be used to create tunable fluids, stabilize mixtures that would otherwise phase separate, and create new materials such as low-fat foods, pastes for printed electronic devices or porous sintered materials~\cite{Koos:2012b, Dittmann:2013, Hoffmann:2014}. 

	\subsection{Capillary suspensions} \label{sec:intro:CapS}

\begin{figure}[tb]
	\centerline{\includegraphics[width=\linewidth]{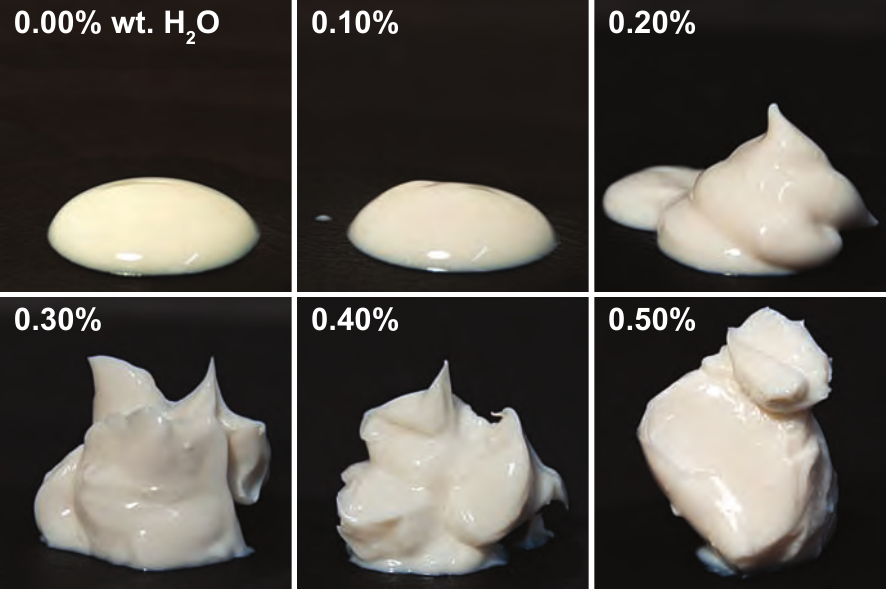}}
	\caption{The transition from weakly elastic, fluid-like behavior to highly elastic, gel-like behavior for capillary state suspension in images. (Hydrophobically modified calcium carbonate, $r = 0.8$~{\textmu}m, in DINP, $\phi = 0.11$, with added water.)\\ Adapted from \cite{Koos:2011}.} 
	\label{fig:socal_pics}
	\end{figure}

Capillary suspensions have been divided into two distinct states: the \emph{pendular state} where the minority liquid preferentially wets the particles; and the \emph{capillary state} where the secondary fluid wets the particles less well than the primary fluid.  Both states are associated with a transition in the suspension from a fluid-like to gel-like state or from a weak gel to a strong gel.    The texture and flow of these suspensions dramatically alter upon the addition of an immiscible secondary liquid at low volume fractions as shown in \autoref{fig:socal_pics}.
In the pendular state, the particle network is formed through individual particles linked through capillary bridges and in the capillary state, clusters of particles surrounding secondary fluid droplets are linked together.  

The creation of the network, formed by adding a preferentially wetting liquid to a suspension, has been investigated previously in higher volume fraction suspensions.  Early work on sedimentation determined that the addition of a secondary fluid (usually water to an organic solvent) caused the particles to flocculate and significantly increased the sedimentation volume~\cite{Bloomquist:1940}. Later, the rheological properties of the suspensions were measured where it was observed that the admixture creates a strong gel and that the yield stress greatly increases~\cite{Kao:1975b}. This network structure was attributed to the formation of pendular bridges between the particles~\cite{Strauch:2012}, though an increase in electrostatic charge on the particle surface might have contributed to some of the observed rheological changes~\cite{Howe:1955}.  An increase in the viscosity for lower volume fractions was also observed by McCulfor and coworkers~\cite{McCulfor:2010}. Rheological changes due to the addition of water to an oil-based suspension of hydrophobically modified calcium carbonate (capillary state) were first observed by Cavalier and Larch\'{e}~\cite{Cavalier:2002}, but they attributed the gelation to hydrogen bonding.  This initial work on pendular state suspensions was extended and the first confirmation of the influence of the capillary force in capillary state suspensions was demonstrated by Koos and Willenbacher~\cite{Koos:2011} where both states were termed capillary suspensions. 

\begin{figure}[tb]
	\includegraphics[width=\linewidth]{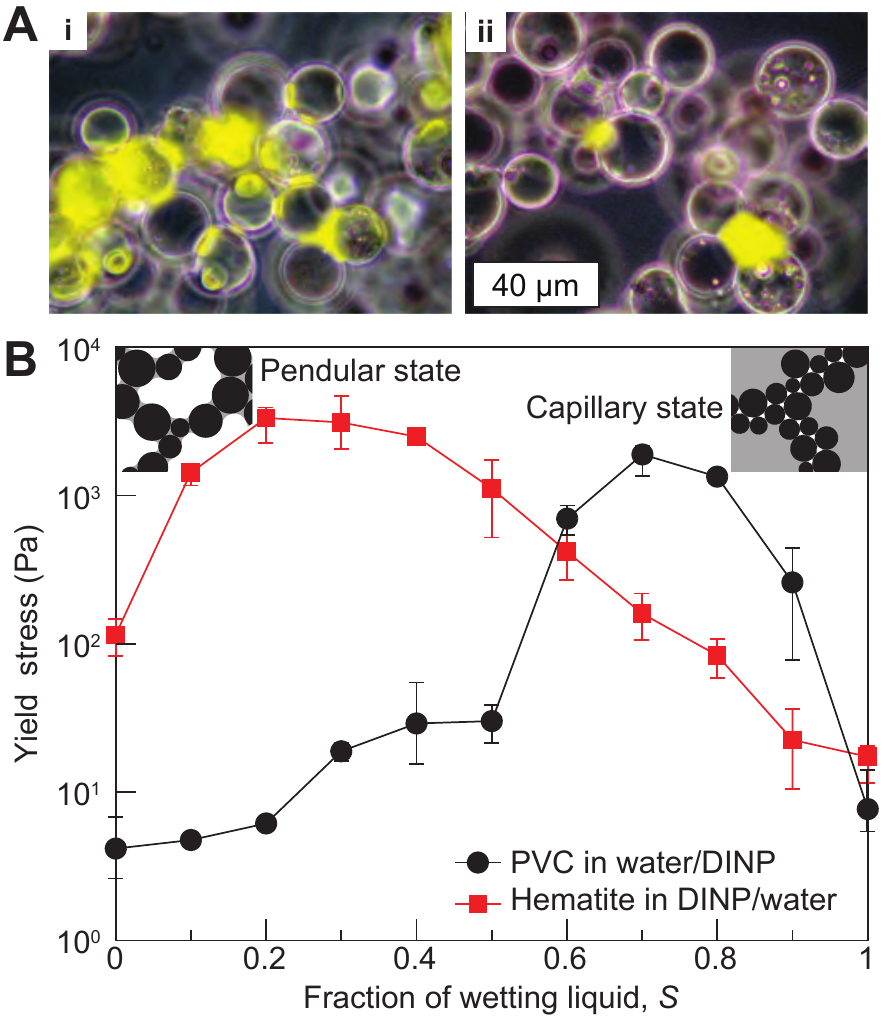}
	\caption{({\bf A}) Images of the (i) pendular and (ii) capillary states. (A.i: Clean glass, $\theta= 49.4^\circ$, and A.ii treated glass, $\theta= 99.3^\circ$, in DINP with 1 wt \% added water. For both samples, $r=12.3$~{\textmu}m and $\phi = 0.3$.) ({\bf B}) Dependence of the yield stress on the fraction of the wetting liquid.  For this PVC, the yield stress is greatest in the capillary state. For hematite, the yield stress is greatest in the pendular state.\\ Adapted from \cite{Koos:2011}.}
	\label{fig:states}
	\end{figure}

These capillary suspensions are differentiated based on which fluid preferentially wets the particles.  Using a fluorescent dye to mark the location of the secondary fluid, a diluted capillary suspension sample may be imaged, as shown in Fig.~\ref{fig:states}A. 
The addition of water to these oil-based suspensions of glass particles causes the particles to agglomerate tightly into clusters.  In the pendular state (Fig.~\ref{fig:states}A.i), the capillary bridges are clearly visible binding particle dimers and trimers together.  A shielding of the second fluid phase by particle clusters is observed in the capillary state (Fig.~\ref{fig:states}A.ii). In contrast, the glass particles without added water agglomerate slightly due to vdW attraction, but are generally well distributed throughout the sample.  

Following the convention of wet granular materials, each sample was characterized using the saturation $S$, 
	\begin{equation}
 	S = \frac{V_\textrm{wetting fluid}}{V_\textrm{total fluid}} = \frac{V_w}{V_b+V_l}
	\end{equation}
which is close to zero for the pendular state ($V_w = V_l$) and approaches one for the capillary state ($V_w = V_b$) . The gel-like transition in capillary suspensions dramatically increases the yield stress and viscosity above the corresponding values of the single-fluid suspensions.  The full dependence of yield stress on the saturation is shown in Fig.~\ref{fig:states}B for PVC ($r=16.4$~{\textmu}m) and hematite, both in various percentages of water and diisononyl phthalate (DINP).  For PVC and similar systems, the greatest yield stress occurs in the capillary state at a fraction of $S\approx0.7$, but for systems similar to hematite (for instance, silica and hydrophilic glass), the yield stress was the greatest in the pendular region ($S\approx0.2$).  While the formation of capillary suspensions is a general phenomenon, each material combination appears to have a preference for either the capillary state or the pendular state.  While this may be a physical restriction depending on, for example, how well the secondary fluid can re-wet the particle surface, it may also depend on the droplet breakup as the suspension is created.

\begin{figure*}[tb!]
	\includegraphics[width=\linewidth]{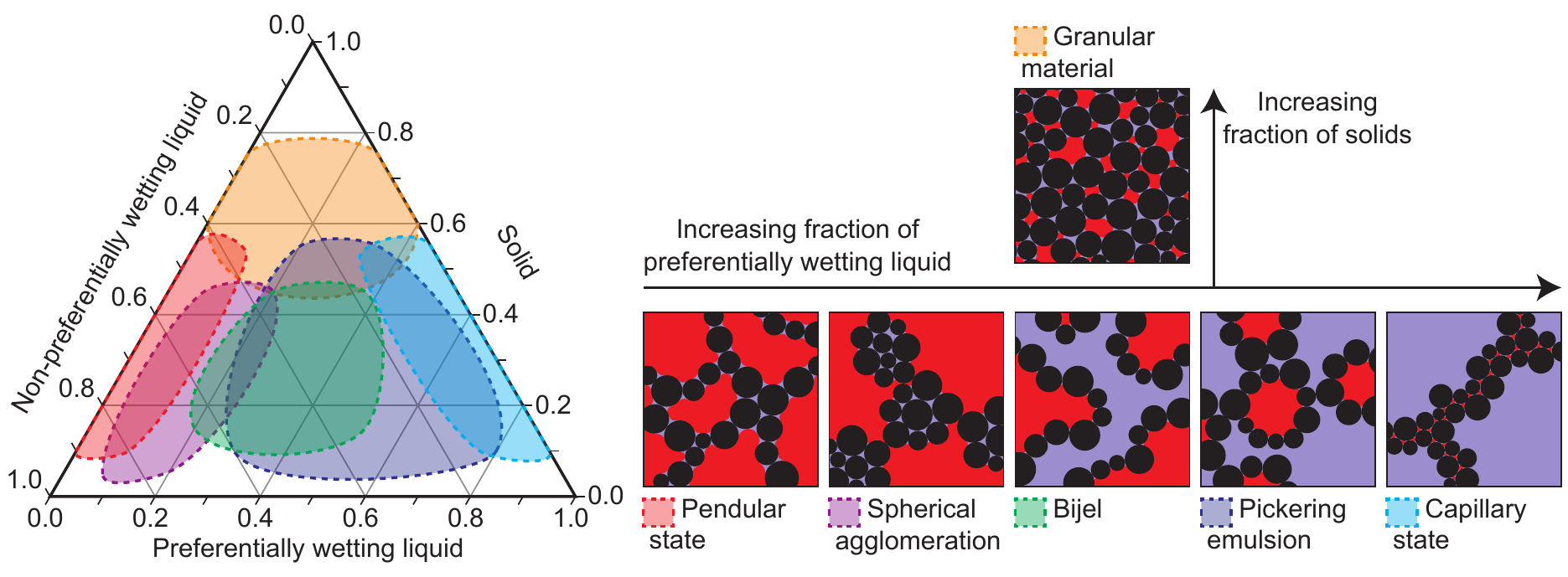}
	\caption{Ternary diagram of particle-liquid-liquid systems showing approximate regions of stability as a function of the relative volume fractions along with the corresponding diagrams for each state. Capillary suspensions occupy the edges where one of the secondary liquids is present as a minority phase at a wide range of particle volume fractions.}
	\label{fig:map}
	\end{figure*}

The pendular and capillary states are shown on a ternary diagram with other particle-liquid-liquid systems in \autoref{fig:map}.
In capillary suspensions, the secondary fluid droplets are typically smaller than the particles and the total secondary fluid volume is a small fraction of the total volume.  In Pickering emulsions, solid particles stabilize the emulsion droplets; hydrophilic particles tend to form stable oil-in-water (o/w) emulsions and hydrophobic particles tend to form water-in-oil (w/o) emulsions.  To create stable Pickering emulsions, the particle size must be much smaller than the droplet size.  The opposite behavior is demonstrated in spherical agglomerates where many particles are enveloped by a larger droplet of preferentially wetting liquid.  Gelation in Pickering emulsions and from spherical agglomerates is caused primarily through the van der Waals interaction among the particles on the surface of adjacent droplets.  The capillary force causes gelation in capillary suspensions. The states available for ternary systems depends on the specific material properties and on the processing history.  Key among these material properties are the interfacial tension and contact angle, both of which contribute to the strength of the capillary force. 

	\subsection{Capillary force} \label{sec:intro:Fc}

When a liquid meniscus forms between two surfaces, an attractive force between these surfaces results.  In suspensions, this capillary force usually dominates over other forces, such as the van der Waals force~\cite{Butt:2009}.  The capillary force, which plays an important role in a wide range of natural phenomena and technical processes~\cite{Reis:2010, Duan:2010}, is composed of two parts:  the Laplace pressure inside the liquid and the surface tension acting at the solid-liquid-gas contact line.  The capillary force $F_c$ between two solid particles connected by a pendular bridge depends on the radius $r$ of the particles, their separation distance $s$, the surface tension of the liquid $\Gamma$, the wetting angle $\theta$, as well as the volume $V_l$ and shape of the liquid bridge.  Analytical as well as computational solutions for $F_c$ assume a certain bridge shape (e.g. toroidal, cylindrical, etc.) or solve the Laplace-Young equation.  For a finite particle separation of equally sized spheres connected by a fluid bridge, the capillary force is given by
	\begin{equation}
	F_c = \frac{2 \pi r \Gamma \cos \theta}{1 + 1.05 \hat{s} + 2.5 \hat{s}^2}, \quad 
	\textrm{with } \hat{s} = s\sqrt{\frac{r}{V_l}}
	\label{eqn:cap:s}
	\end{equation}
which simplifies to the well-known expression $F_c = 2 \pi r \Gamma \cos \theta$ for spheres that are in contact~\cite{Herminghaus:2005}.  The equations for the capillary force may be modified to account for spheres of different sizes~\cite{Willett:2000} or surface roughness~\cite{Butt:2008}.

The relationship between the capillary force connecting individual grains and the macroscopic stress for a sample depends on the coordination number, i.e. the number of contacts per particle in the agglomerate and the number of particles per unit area.  For a uniform packing of equally sized spheres with liquid bridges between particles in direct contact, the relationship is given by,
	\begin{equation}
	\sigma_y = f(\phi) \frac{F_c}{r^2} = f(\phi) \frac{2 \pi \Gamma \cos\theta}{r}
	\label{eqn:yield}
	\end{equation}
where $f(\phi)$ is a function of the particle volume fraction.  For $\phi \ll 1$, as is often the case in capillary suspensions, $f(\phi) = \phi^2$ would be a reasonable approximation.

The very strong capillary force may lead to distinct flow behavior in suspensions.  Indeed, the gel strength or yield stress is orders of magnitude higher than with van der Waals attraction and can be tuned in a wide range through the amount of added secondary liquid.  Typically, the capillary force $F_c$ is two or three orders of magnitude stronger than the van der Waals force $F_{vdW}$. The ratio $F_{vdW}/F_c$ is independent of particle size in first approximation.

\section{Particle clusters} \label{sec:cluster}

	\subsection{Colloidal clusters} \label{sec:cluster:Pick}
\begin{figure*}[tb]
	\centerline{\includegraphics[width=\linewidth]{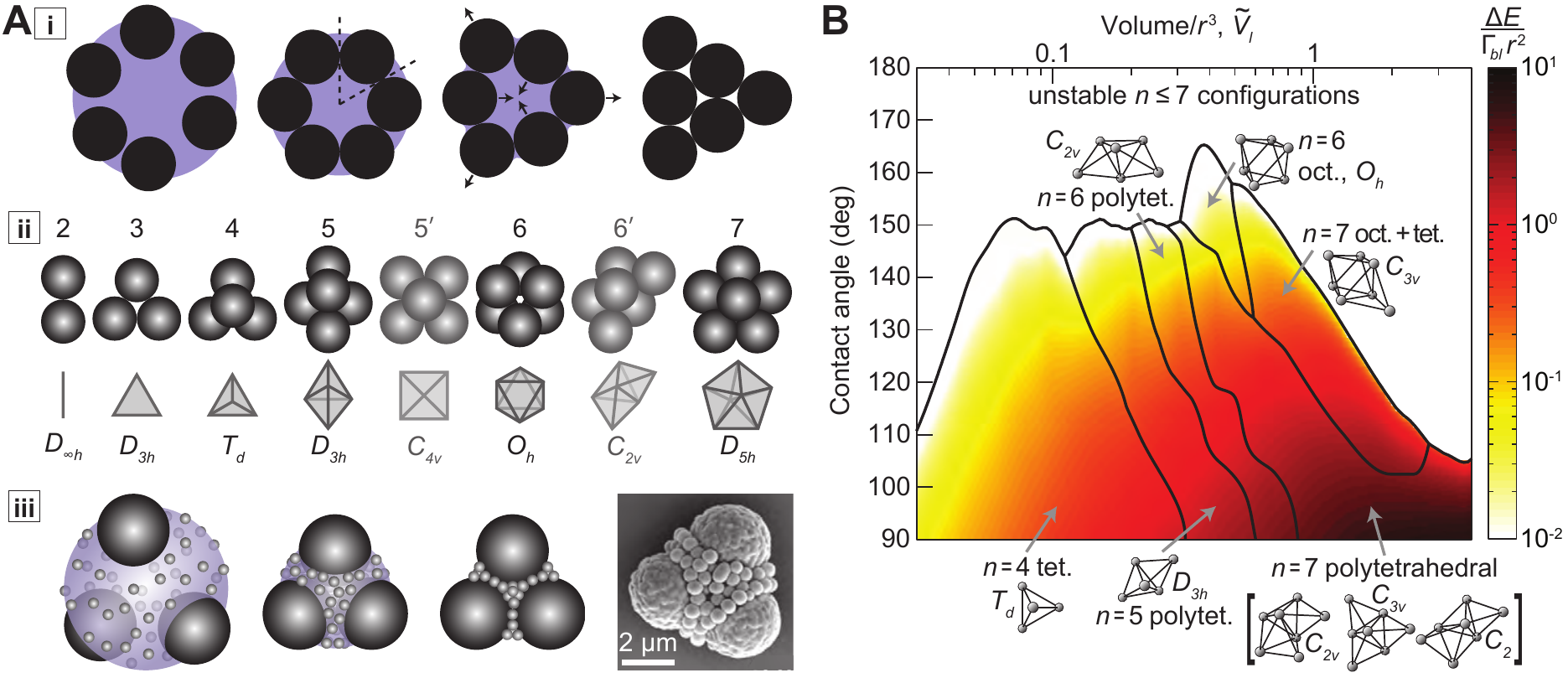}}
	\caption{({\bf A}) Colloidal clusters from emulsions showing (i) their formation, (ii) Minimum second moment configurations with alternative 5' and 6' configurations, and (ii) clusters with bimodal particle sizes. ({\bf B}) State diagram showing the minimum energy regions for various particle clusters as a function of the normalized secondary fluid droplet volume (normalized by the particle radius, $r$, cubed) and the wetting angle $\theta$. Color indicates the reduction in energy of respective clusters over the configuration of individual detached fluid drops and particles.\\ Adapted from~\cite{Lauga:2004, Manoharan:2003, Cho:2008, Koos:2012}.} 
	\label{fig:cluster}
	\end{figure*}

	Evaporating emulsion droplets when small numbers of particles are attached to these droplets can create clusters of particles. Such \emph{colloidal clusters} were described by Manoharan et al.~\cite{Manoharan:2003}, where it was hypothesized that these clusters form structures that minimize the second moment of mass distribution.  This reorganization of the droplets into clusters was modeled by Lauga and Brenner~\cite{Lauga:2004} and is shown in \autoref{fig:cluster}A.i.
The particles follow the interface until they reach close-packing on the droplet surface.  At this point, the capillary force and contact forces will cause the particles to pack into tight clusters with a single configuration for each number of particles, $n$.  These configurations are shown in \autoref{fig:cluster}A.ii for clusters between $n=2$ and $n=7$ particles.  These minimum second moment configurations assume that the evaporation of the secondary fluid proceeds sufficiently slow so that the particles can rearrange.  Chirality can be introduced into each cluster configuration by using dumbbell shaped particles~\cite{Peng:2013}. 

One problem with creating colloidal clusters from emulsion droplets is that droplet polydispersity creates a distribution of cluster sizes~\cite{Brugarolas:2013}.  This distribution of particles in colloidal clusters and their applications has been studied by the groups of Pine and Manoharan. These groups recently found that this distribution in sizes is improved using microfluidic methods such as the rotating cup, which creates clusters with a narrow distribution in particle numbers~\cite{Yi:2002} or optimized shearing conditions~\cite{Wang:2012}. The Wittemann group demonstrated that it is also possible to produce these clusters from miniemulsions using ultrasonication~\cite{Wagner:2008, Wagner:2010, Wagner:2012}.  

Methods designed to produce predominantly small clusters can also produce non-minimum second moment isomers such as the $n=5$ square pyramid (5' in \autoref{fig:cluster}A.ii, $C_{4v}$ configuration)~\cite{Wagner:2010}.  These were modeled by Schwarz and coworkers~\cite{Schwarz:2011} using Monte Carlo simulations where the colloidal particles had a short-ranged attraction and long-ranged repulsion potential. They attributed the presence of non-minimal isomers to the attractive potential between particles and found that while two different potentials tested produce the same distribution of particle numbers in the clusters, the ratio of isomers varied for each $n$~\cite{Schwarz:2011}.  Furthermore, the attractive potential between clusters allowed for superclusters (hierarchical assembly of smaller clusters) to form.  It is interesting to note that clusters can also be produced using a sparse suspension of spherical particles with an attractive potential alone, as shown by the Brenner group~\cite{Arkus:2009, Meng:2010}.  Using the attractive potential (e.g. through a depletion interaction) alone will produce clusters that are less compact, that is with lower bond numbers, due to the increase in the number of possible routes for cluster formation.  For example, a polytetrahedral $n=6$ cluster (6' in \autoref{fig:cluster}A.ii, $C_{2v}$) will be highly favored over the octahedral structure ($O_{h}$)~\cite{Meng:2010}.  Thus, the computational work by Schwarz can be viewed as occupying an intermediate state in the transition between the compact clusters formed by evaporating Pickering emulsions and the irregular clusters created using an attractive potential where clusters can be kinetically arrested during evaporative-driven rearrangement.  

Creating clusters from Pickering emulsions using bidisperse particle mixtures was the focus of recent work by the group of Yang~\cite{Cho:2005, Cho:2008}.  As the emulsion droplet dries, the arrangement of large particles will proceed as before: reaching first a close-packed configuration on the droplet surface and then transforming into a minimum moment structure.  The small particles will continue to follow the interface, as shown in \autoref{fig:cluster}A.iii, reinforcing the junctions between adjacent particles~\cite{Cho:2008}.  The coverage of the large particles will vary depending on size and volume ratios of the particles as well the interparticle and interphase interactions, with smooth colloidal aggregates created when the secondary particles are much smaller than the large particles~\cite{Cho:2008}.  Clusters from a single material (e.g. silica) can be sintered together to form a new colloidal particle with complex structure. Disparate materials can be used to create Janus composite clusters or hollow clusters~\cite{Cho:2008}.  Such composite clusters can be mimicked using a photocurable material as the secondary fluid~\cite{Kim:2008}.  

These colloidal clusters can be functionalized to control their arrangement into large self-assembled structures including crystals. These crystalline structures can be useful for photonic devices, particularly if crystal structures possessing large photonic band gaps can be easily and consistently manufactured.  Patchy clusters, e.g. clusters from heterogeneous large and small particles, with amphiphilic properties can be used to form colloidal molecules~\cite{Wang:2012, Kraft:2012} and lattice structures~\cite{Mao:2012}.  Using an applied electric field, Forster and co-workers were able to form an ordered crystal structure~\cite{Forster:2011}.  Wang et al. also showed that it is possible to use colloidal clusters to form lock and key particles, which can also be used to assemble colloidal molecules~\cite{Wang:2014b}.

	\subsection{Model for capillary suspensions} \label{sec:cluster:model}
	
Capillary state suspensions are composed of clusters of particles shielding the small secondary fluid droplets, which are hypothesized to form the building blocks for the sample-spanning network.  Using the previous work on colloidal clusters as a guide, the clusters forming the building blocks of capillary suspensions can be modeled.  These clusters contain few particles arranged in tetrahedral or octahedral arrangements~\cite{Koos:2012}.  These clusters were modeled using the computational code Surface Evolver designed by Brakke~\cite{Brakke:1992} and the result of this model is shown in \autoref{fig:cluster}B. The lowest energy states are presented  as a function of the contact angle $\theta$ and normalized droplet volume $\tilde{V}_l$.  Nine different particle configurations were studied representing all possible close-packed structures using between 4 and 7 particles using both new seed groupings as well as structures composed of smaller seeds.  This computation model found that structures composed of tetrahedral clusters (each surrounding an individual droplet) had the greatest reduction of energy for contact angles $\theta_c < 151.2^\circ$ for lower drop volumes. At higher contents of secondary fluid and contact angles $\theta_c < 165.3^\circ$, octahedral and mixed octahedral-tetrahedral clusters will be stable~\cite{Koos:2012}.  
 
This computation model of Koos and Willenbacher correctly predicted the rheological behavior of an oil-based capillary state suspension~\cite{Koos:2012}.  The admixture will begin to form tetrahedral clusters with low added water content, but this does not result in a significant rise in the yield stress.  With additional water, the admixture will form a percolated tetrahedral network and then transition to an octahedral structure.  The formation of a tetrahedral network results in a rapid increase in the yield stress while the transition to an octahedral network causes only a modest additional increase~\cite{Koos:2012}.  

\section{Rheological properties} \label{sec:rheo}

The addition of a secondary fluid can induce a transition from fluid-like to gel-like behavior or from a weak gel to a strong gel due to the formation of a strong network induced by the capillary force.  This transition is shown in sample images (\autoref{fig:socal_pics}) and in the magnitude of the shear modulus $|G^{\ast}|$ as a function of the oscillatory frequency for a CaCO$_3$ suspension in an organic solvent. The shear modulus transitions from a linear dependence on the frequency -- indicative of fluid-like systems -- without added water to frequency invariance -- denoting a gel-like system -- at a water content of 0.20\% for this suspension.  At higher water contents, the shear modulus remains frequency independent and increases in magnitude. The formation of the sample-spanning network in capillary suspensions is based, in part, on the volume fraction of solids.  For sparse suspensions, the addition of a secondary liquid will cause the formation of flocs.  These flocs can either remain stable -- if the suspension is otherwise well stabilized and the density mismatch is minimal -- or more commonly, these flocs will settle under gravity to produce a 1-phase region and a 3-phase region~\cite{Hijnen:2014}.  At the percolation point, the capillary suspension will form a sample-spanning network.  If the network is weak, it cannot sustain its own weight at this point and a supernatant will develop on the admixture if left for long times.

\begin{figure*}[tb]
	\includegraphics[width=\linewidth]{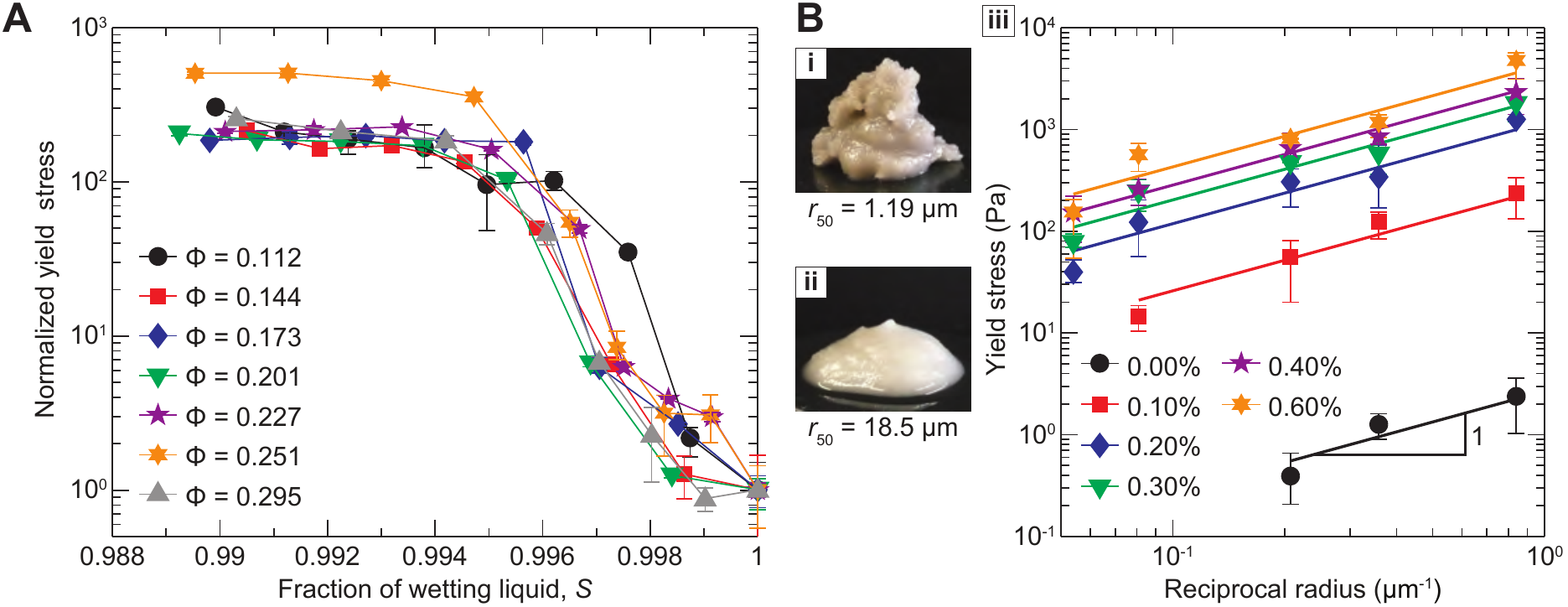}
	\caption{({\bf A}) Yield stress, normalized by the stress without added secondary fluid ($S=1$), for various solid fractions. ({\bf B}) Images of (i) $r=1.19$ {\textmu}m and (ii) $r=18.5$ {\textmu}m sized particles both with 0.6\% added water as well as (iii) yield stress as a function of the particle reciprocal radius. 
(A: CaCO$_3$ $\phi=0.173$ in diisononyl phthalate with added water, capillary state, B: glass $\phi=0.30$ in diisononyl phthalate with added water, pendular state, from~\cite{Koos:2011, Koos:2012b}.)}  
	\label{fig:rheo}
	\end{figure*}

The influence of the particle volume fraction above the percolation threshold has been investigated.  Admixtures of hydrophobically modified CaCO$_3$ in DINP with volume fractions between $\phi = 0.112$ and $\phi = 0.295$ were used. The amount of water (the secondary fluid) was varied between 0.0 and 0.8\% by mass, corresponding to a saturation $S$ between 1.00 and 0.99. Each mixture was characterized by measuring the yield stress, as shown in Fig.~\ref{fig:rheo}A. 
This yield stress is normalized against the yield stress measured when no water is present ($S=1$) and is displayed as a function of the wetting-liquid fraction.   Addition of water increases the yield stress by more than two orders of magnitude over the single-fluid suspensions.  This increase in the normalized yield stress is independent of the volume fraction of solids~\cite{Koos:2011}. The characteristic curve shape indicates that the most rapid change in yield stress occurs at a saturation of 0.997.  Below $S=0.994$, the curves plateau.  As this is a capillary state suspension, we can use the model discussed in \autoref{sec:cluster:model} to predict the transitions based on the particle subgroupings.  The initial increase in the yield stress corresponds to the formation of a tetrahedral network (rather than tetrahedral flocs) and the plateau or minimal increase in strength corresponds to the transition into subgroupings of higher order~\cite{Koos:2012}. In the pendular state, the dependence of the yield stress on the solid volume fraction exhibits nearly identical behavior~\cite{Dittmann:2013}. At higher secondary fluid contents, however, the samples exhibit a peak yield stress (likely denoting the transition from a capillary network to agglomerates). The location of this peak depends on the solid fraction with higher solid fraction samples having a peak yield stress at higher secondary fluid contents~\cite{Dittmann:2013}.
	
	\subsection{Tunable strength} \label{sec:rheo:tuning}
		
The strength of these ternary admixtures can be modified by changing the capillary force joining individual particles.  The capillary force and the corresponding yield stress of the pendular state network are given by \autoref{eqn:yield} where the yield stress should be proportional to the interfacial tension and reciprocal particle radius.  The interfacial tension can be modified through the choice of the two fluids, as can be shown using aqueous glycerol of varying concentrations as the secondary fluid.  A linear dependence on the interfacial tension is apparent for each solid volume fraction tested.  The interfacial tension can also be modified by changing the sample temperature, a route that can be used to externally tune the strength of capillary suspensions in situ~\cite{Koos:2012b}.  In certain capillary suspensions, it is also possible to remove the secondary fluid returning the suspension to its original state.  This reversible process was demonstrated for a CaCO$_3$ suspension in silicone oil with water added and removed several times with no apparent degradation of the materials~\cite{Koos:2012b}.

A strong dependence of network strength on the particle size was expected for these admixtures due to the variation in capillary force and area of contact. This influence of size variation is apparent in \autoref{fig:rheo}B for a pendular state suspension.  The particle admixtures created with smaller particles (\autoref{fig:rheo}B.i) retain a fairly rigid shape whereas the admixtures with larger particles spread under their own weight (\autoref{fig:rheo}B.ii) clearly showing the difference in yield stress between these two mixtures.  \autoref{fig:rheo}B.iii shows that the $r^{-1}$ dependence, predicted by  \autoref{eqn:yield}, is clearly obtained for all fractions of added water shown~\cite{Koos:2012b}.  The viscosity, at a wide range of shear rates, also increases with decreasing particle size~\cite{McCulfor:2010}. While using smaller particles will produce stronger capillary suspensions, they also require smaller secondary fluid droplets to be produced during mixing.  Agglomerates or Pickering emulsions will be created if the droplets are larger than the particle decreasing the strength of the ternary mixture. This limitation on the droplet size may make the use of nanoparticles in capillary suspensions difficult.
		
The dependence on particle size and interfacial tension both confirm the relevance of the capillary force.  These variations also give insight into the tunability of these mixtures.  Additional experiments with surfactants also show that increasing surfactant concentration decreases the strength of capillary suspensions~\cite{Koos:2012b}.  The relationship between the yield stress and the surfactant concentration depends on the surfactant type, but no correlation with the hydrophilic-lipophilic balance is demonstrated. This reduction is greater than the change in interfacial tension as the surfactant decreases the number of capillary bridges by independently stabilizing emulsion droplets~\cite{Koos:2012b}. Thus, by adding surfactants the strength of the capillary suspensions can be decreased as desired or the capillary bridges can be prevented, returning the admixture properties to the condition without added fluid (albeit in an irreversible manner).

	\subsection{Mixing conditions} \label{sec:model:mixing}
	
\begin{figure}[tb]
	\includegraphics[width=\linewidth]{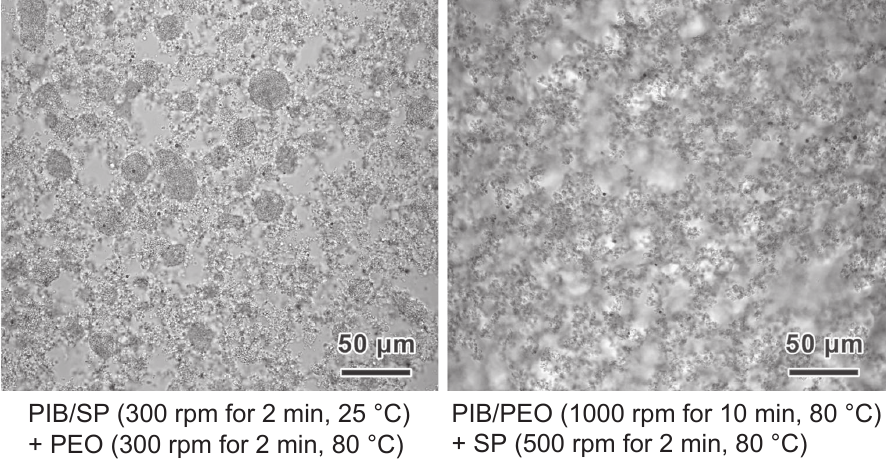}
	\caption{Confocal micrographs (DIC) of PIB/PEO/SP ternary blends obtained for two mixing procedures (blend of 77\% polyisobutylene (PIB), 20\% spherical silica particles (SP), and 3\% polyethylene oxide (PEO), pendular state, from~\cite{Domenech:2014}).}  
	\label{fig:mixing}
	\end{figure}

Changing the material properties and ratios as well as the mixing conditions can modify the microstructure of capillary suspensions. This latter factor was investigated in two recent papers by the Velankar group using particles that were preferentially wetted by the secondary fluid~\cite{Heidlebaugh:2014, Domenech:2014}. These particles can migrate into large fluid droplets during mixing forming spherical agglomerates.  As the droplet becomes saturated with particles, its effective viscosity increases many times making it virtually impossible to break apart.  While these aggregates are individually strong, they are detrimental to the formation of a strong sample-spanning network as they increase the total solid fraction necessary for percolation and the weak link between aggregates can be easily broken.  The size of these aggregates created during mixing should depend on the relative rate of break-up or coalescence as well as the rate of addition and removal of particles to the droplet. To prevent these aggregates, the suspension is mixed at a very high initial speed such that large droplets are broken apart into sufficiently small droplets before they can become saturated with particles. Domenech and Velankar showed that low intensity mixing resulted in highly polydisperse drop sizes while intense mixing created droplets that were smaller and had a narrower distribution in sizes~\cite{Domenech:2014}.  A pendular network was created under both mixing conditions (as the smallest droplet size was comparable to or smaller than the particle size in both cases), but the network strength was significantly enhanced with strong mixing~\cite{Domenech:2014}. Alternatively, the capillary suspension can be created in the other order where the particles are added to an emulsion.  While the secondary fluid droplets can rapidly coalesce and separate, this method creates very uniform capillary suspensions without aggregates if they are mixed into the emulsion sufficiently fast as shown in \autoref{fig:mixing} for an immiscible polymer blend with spherical silica particles~\cite{Heidlebaugh:2014, Domenech:2014}.  
The uniform droplet sizesizes formed during the high-speed mixing shown in \autoref{fig:mixing}B creates a much stronger capillary suspension with a viscosity that is two times higher for an immiscible polymer blend with spherical silica particles than for the less intense mixing (\autoref{fig:mixing}A).  

Even if aggregate formation is avoided during mixing, aggregation will still occur at a sufficiently high content of secondary wetting fluid. In Heidlebaugh et al.~\cite{Heidlebaugh:2014}, the Velankar group hypothesized that aggregation will be initiated when a capillary meniscus can be shared by three or more particles.  Any applied flow or Brownian motion will transition these binary bridges into ternary or higher aggregates.  By assuming a monodisperse distribution of particles and bridges, the critical secondary fluid content $S_c$ can be calculated as
\begin{equation}
S_c =\frac{3}{8\pi} \tilde{V}_{l,c} \frac{\phi}{1-\phi} z
\end{equation}
where $ \tilde{V}_{l,c}$ is the critical binary bridge volume, which depends on the wetting angle, normalized by the particle radius cubed and $z$ is the number of menisci per particle. The transition to a cluster of three particles is estimated as $S_c \approx 0.054 z \phi/(1-\phi)$
for particles in contact where the initial bridge was small and cylindrical~\cite{Heidlebaugh:2014}. The transition from a majority of binary bridges to ternary clusters in a pendular state suspension will correspond to a decrease in the strength of this network. While Heidlebaugh and coworkers found qualitatively good agreement with their experiments for $z=4$, their choice of the coordination number was arbitrary. If this analysis is modified to use a toroidal approximation with a contact angle $\theta = 0^\circ$, a smaller transitional value of $S_c \approx 0.024 z \phi/(1-\phi)$ is calculated.  For a volume fraction of $\phi = 0.1$ and coordination number of 4, the transition to ternary clusters should proceed at a secondary fluid content $S=1\%$ -- lower than is usually observed in experimental systems. Domenech and Velankar found good agreement with their data using the toroidal approximation and a coordination number of $z=2$, finding a peak yield stress at $\phi_l/\phi = 0.22$, where $\phi_l$ is the volume fraction of secondary fluid~\cite{Domenech:2015}. The formation of larger clusters with six particles (octahedral structure) is predicted to occur at $\phi_l/\phi = 0.53$, which corresponds to a steep reduction in the measured yield stress~\cite{Domenech:2015}. Polydisperse bridges and particles should further shift the location of peak pendular suspension strength to lower saturations. The transition to clusters occurs at slightly higher saturations when intermediate contact angles are used with a maximum at $S_c \approx 0.071 z \phi/(1-\phi)$ for $\theta = 68.5^\circ$ corresponding to 2.4\% added fluid for $\phi = 0.1$ and $z=4$. 

In capillary state suspensions, where the secondary fluid does not preferentially wet the particles, inadequate mixing will also detrimentally affect the material strength.  Large droplets will become rapidly coated with particles creating Pickering emulsions.  The size of these droplets decreases with increasing particle concentration for the same mixing conditions due to stabilization of the small droplets by the particles~\cite{Aveyard:2003}, but as the droplet size approaches the particle size, local flow fluctuations induced by the particles dominate the droplet break-up leading to increased polydispersity~\cite{Kaur:2010}.

	\subsection{Network structure} \label{sec:rheo:network}
	
In the capillary state, the clusters of particles surrounding a secondary fluid droplet must aggregate and form a sample-spanning network. These clusters can rearrange depending on the strength of attraction and externally applied forces. Using small amplitude oscillatory shear measurements, it has been shown that the end of the linear viscoelastic region precedes the crossover amplitude by orders of magnitude~\cite{Koos:2014}.  This broad transition regime between the onset of non-linear response and yielding is likely due to the decreasing strength of the capillary bridges with particle separation before rupture.  Variations in droplet size and cluster configurations further intensify this effect as small bridges or weak clusters can rupture earlier than large bridges and strong clusters. The sensitivity of the suspensions to oscillatory deformations means that care must be taken when measuring the viscoelastic properties of capillary suspensions, particularly in the capillary state.

In order to study the formation and aging of the capillary suspension, the storage $G'$ and loss $G''$ moduli can be monitored as a function of time since rejuvenation.   At low amplitudes, the aging follows a weak power-law dependence~\cite{Koos:2014}. This power-law aging persists for very long times -- several days -- with no apparent change in the slope.  The aging in quiescence was shown to differ from aging with continuous oscillation in one experiment~\cite{Koos:2012}, but this is likely due to the oscillatory amplitude used.  The behavior of the weak network formed by the admixture with 0.10 wt.\% water is distinctly different  progressing more rapidly. In this case, $G''$ initially dominates over $G'$. Both moduli increase with time and after a short period $G'$ is larger than $G''$ indicating that under this light oscillation, the initially weak network is able to rearrange into a much stronger state~\cite{Koos:2012, Koos:2014}.  The cluster model (\autoref{sec:cluster:model}) for this particular admixture predicted a fragile percolated network consisting of tetrahedral subgroups as well as free particles -- a weak network that should be easily reorganized to increase the network strength.

Higher oscillatory amplitudes can enhance the rearrangement of these capillary suspension networks, but also cause their rupture.  At low strain amplitudes (e.g. $\gamma = 0.1\%$), the network increases continually for the experimental duration~\cite{Koos:2014}.  At higher strains, $G'$ initially increases in strength (while $G''$ remains fairly constant) until a specific time at which both $G'$ and $G''$ decrease suddenly never to recover.  This behavior becomes more dramatic at higher strain amplitudes where no strengthening is observed and $G''>G'$ throughout the experiment~\cite{Koos:2014}. This behavior differs from the suspensions without added water where increasing the oscillatory amplitude did not result in a sudden rupture of the network.  This rearrangement and fracture indicates that capillary suspensions can be sensitive to applied forces and the time since their creation.

\begin{figure}[tb]
	\includegraphics[width=\linewidth]{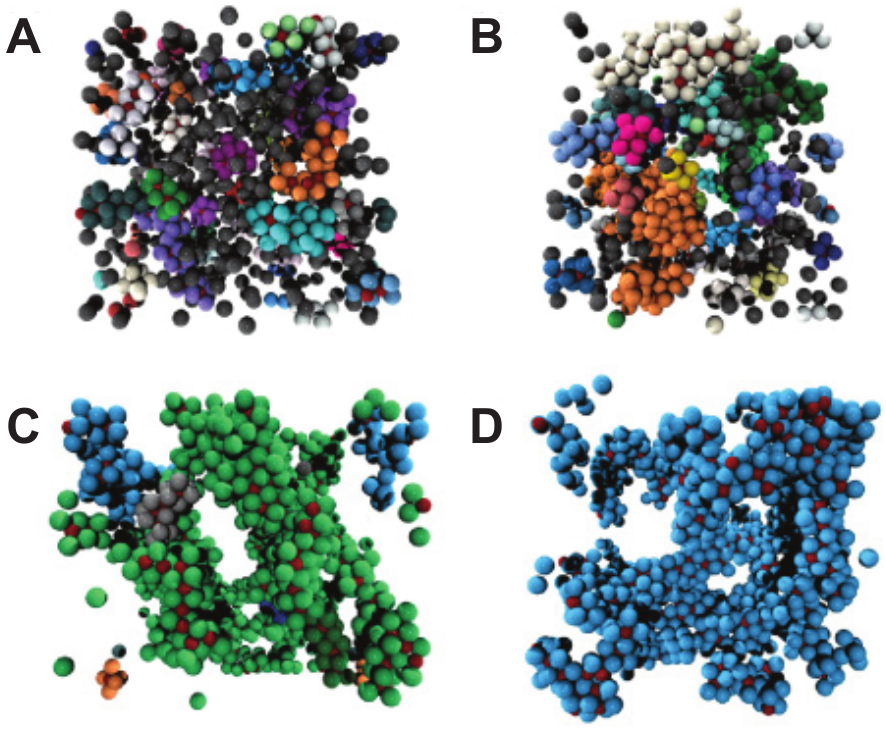}
	\caption{Simulation snapshots for particles aggregated by droplets for ({\bf A}) Fluid of clusters at $S=0.70\%$, ({\bf B}) fluid of clusters at $S=1.17\%$, ({\bf C}) fluid at percolation point $S=1.64\%$, and ({\bf D}) $S=2.2\%$ percolated state. Droplets are shown as small red spheres while each individual cluster is given a unique color. (Taken at simulation time $t=800 (2r)^2/D_0$, where $D_0$ is the Stokes-Einstein diffusion coefficient and droplet radius $r_l=0.75 r$, from~\cite{Fortini:2012}.)}  
	\label{fig:network}
	\end{figure}

Using Brownian dynamics simulations, Fortini~\cite{Fortini:2012} found that the addition of non-wetting droplets could lead to the formation of clusters and the percolation of these clusters into a sample-spanning network. The formation of clusters was particularly stable when the entire surface of the droplet is covered by particles, confirming the results described in \autoref{sec:cluster:model}.  A sufficient volume fraction of droplets was required before the network would percolate and that the required saturation depended on $\tilde{V}_l$ with larger droplets requiring a greater number of droplets, thus higher secondary fluid contents.  This dependence on the secondary fluid volume fraction $\phi_l$ is shown in \autoref{fig:network} 
for droplets with a radius $r_l$ smaller than the particle radius ($r_l = 0.75 r$) for a specific simulation time.  At low saturations (\autoref{fig:network}A and B), the particles will flocculate, but no sample-spanning network will be formed.  Above a percolation saturation of $S=1.64\%$ (\autoref{fig:network}C and D), the particle clusters will form a sample-spanning network with this network forming earlier for higher water contents (\autoref{fig:network}D). 

This dependence of the formation of a sample-spanning network on droplet size implies that given larger initial droplet sizes, the percolation based solely from tetrahedral clusters, as shown in \autoref{fig:cluster}, is inaccessible shifting the required saturation before percolation higher. Fortini also found that the initial cluster growth was diffusion limited (DLCA) with a fractal dimension $d_f = 1.80 \pm 0.05$.  At larger cluster sizes, the dimensionality was dominated by the random aggregation of different mesoscopic clusters with $d_f = 2.6 \pm 0.1$~\cite{Fortini:2012}. These results have not yet been experimentally verified by direct imaging or scattering data.

In the pendular state, Domenech and Velankar measured the elastic modulus at the plateau $G'_p$ and yield stress $\sigma_y$ as a function of the volume fraction of solids~\cite{Domenech:2015}. For a system consisting of silica spheres in polyisobutylene with added polyethylene oxide, they found a scaling of $G'_p \sim \phi^m$ and $\sigma_y \sim \phi^n$ with $m=4.9$ and $n = 3.3$ for $\phi$ between 0.01 and 0.4.  (These values differ from those obtained from the silica-polyisobutylene pastes of $m=9.4$ and $n=6.1$ for $\phi = 0.2$--0.6.)  Using the scaling law of de Gennes, Domenech and Velankar found $d_f = 1.98$ and $d_f = 1.79$ for the $G'_p$ and $\sigma_y$ data, respectively~\cite{Domenech:2015}. The value obtained from yield stress data is close to that of diffusion limited aggregation ($d_f \approx 1.8$), but the dimensionality from the elastic modulus is between DLCA and reaction limited growth (RLCA, $d_f \approx 2.3$).  While these two methods should agree, these differences and the underlying difference between yield values obtained from strain-controlled and stress-controlled measurements imply that capillary suspensions may have either a heterogeneous network structure or changes in the kinetics near network rupture~\cite{Domenech:2015}.

\section{Spontaneous formation}
\begin{figure}[tb]
	\includegraphics[width=\linewidth]{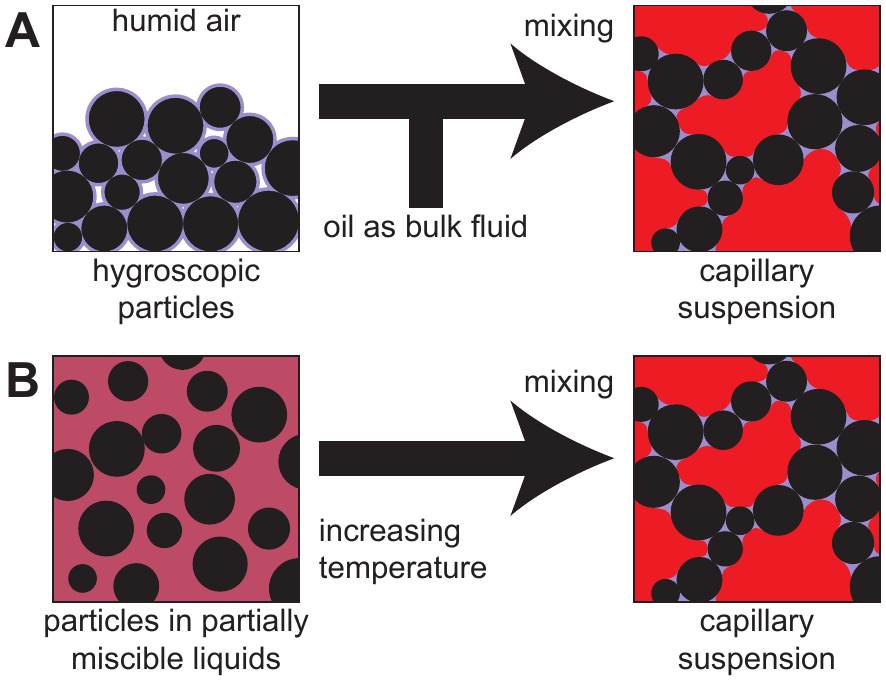}
	\caption{Formation of a sample-spanning capillary network ({\bf A}) upon the addition of oil to hygroscopic particles stored in a humid environment and ({\bf B}) demixing of partially miscible fluids with an increased temperature.}
	\label{fig:spont}
	\end{figure}

The spontaneous formation of a sample-spanning network can occur when particles coated with a small amount of fluid are immersed into an immiscible, bulk fluid or upon a change in temperature in partially miscible fluids, as sketched in \autoref{fig:spont},
and are applicable to capillary state as well as pendular state suspensions.  Using starch stored over water in a closed vessel, Hoffmann and coworkers~\cite{Hoffmann:2014} found that the yield stress of the conditioned starch particles suspended in sunflower oil closely matched the results of dry starch in oil with added water. An energetically unfavorable sharp edge is created between the water layers when the starch granules come into contact during mixing, prompting the formation of a capillary bridge~\cite{Herminghaus:2005}. The contact angle for water on starch is below 90$^{\circ}$ in air, but the three-phase contact angle is 126$^{\circ}$ implying that a capillary state suspension was created~\cite{Hoffmann:2014}. This experiment has been repeated for other hygroscopic particles including cocoa, PVC, and CaCO$_3$ all with similar results.  The suitability of specific particles and their agreement with capillary suspensions created through the addition of water to the pure suspension depends on the particle roughness and the particle porosity~\cite{Hoffmann:2014}.

It is also possible to induce the formation of capillary bridges using phase separating fluids. When the temperature increases above the binodal line, demixing of the solvents is initiated.  The fraction of the secondary fluid available for bridging depends on the ratio of the two fluids in the 1-phase region and the sample temperature.  This was demonstrated in higher volume fractions using water-lutidine by G\"{o}gelein et al. where the bridge size and corresponding capillary cluster size could be controlled~\cite{Gogelein:2010}.  The capillary clusters created from these partially miscible fluids are extremely uniform.  Moreover, the computational work by Cheng and Wang showed that the bridge locations and volumes are exceedingly stable~\cite{Cheng:2012}.  In a cluster of three particles, the menisci form at the location of the closest approach between each particle pairs (beginning with the smallest radius of curvature) and each bridge grows larger until the desired saturation is reached.  These bridges do not spontaneously transition to fill the region between the three particles~\cite{Cheng:2012}. This means that pendular state suspensions created from phase separating liquids will not transition into spherical agglomeration clusters without external forcing.  This model, however, does require that the particles are sufficiently close when demixing occurs or that the particles undergo some Brownian motion.  For sparse suspensions, Hijnen and Clegg demonstrated that shear-induced aggregation can cause phase separation~\cite{Hijnen:2014}.  The return to a stable charge-stabilizied suspension upon sonification implies that the capillary aggregates can be broken apart.  These broken flocs are most likely to be small spherical aggregates rather than individual particles with a secondary fluid layer.  Thus, the suspension cannot be returned to the initial, premixed state without removing the secondary fluid by once again decreasing the temperature to the 1-phase region.

\section{Applications}

Capillary suspensions can aid in our understanding of existing materials, can be used to tune material properties to meet process or application demands, or even to even create new materials.  The addition of a small amount of secondary fluid to a suspension creates a strong sample-spanning network that prevents sedimentation so that these suspensions can be stored for long periods of time without the need for continuous agitation or remixing before use even in the case of a strong density mismatch between the bulk fluid and particles~\cite{Koos:2011}. Additionally, this capillary network can be used to tune the rheological properties of materials by changing the strength of the capillary force or by preventing the capillary interactions through the addition of surfactants~\cite{Koos:2012b}.

\begin{figure*}[tb!]
	\includegraphics[width=\linewidth]{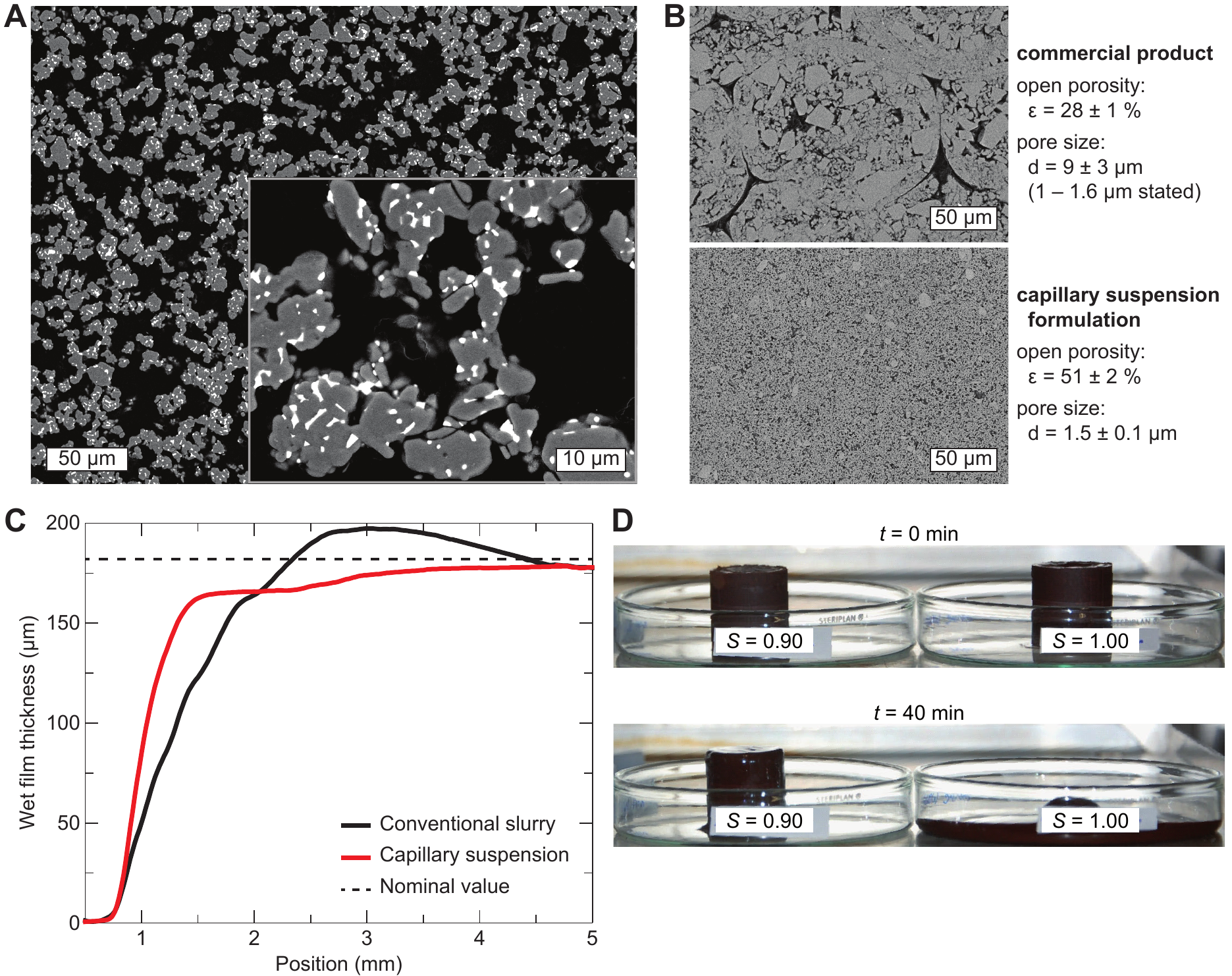}
	\caption{Applications of capillary suspensions. ({\bf A}) SEM slice of porous ceramic where contacts between Al$_2$O$_3$ particles (gray) are reinforced using ZrO$_2$ (white). ({\bf B}) SEM slice of porous glass filters from a capillary suspension formulation and commercial product. ({\bf C}) Edge contour of wet, coated lithium-ion battery electrode slurries. ({\bf D}) Stability of model chocolate during storage at 50 $^\circ$C. (A: $\phi = 0.10$ Al$_2$O$_3$ with $\phi = 0.05$ ZrO$_2$ in paraffin oil with added aqueous sucrose solution, sintered at 1650 $^\circ$C for 2 h, B: SiO$_2$ in paraffin oil with added aqueous sucrose, C: conventional paste $\phi = 0.20$, $S = 0.00$ and capillary suspension slurry $\phi= 0.20$, $S=0.02$, and D: cocoa particles in cocoa butter $\phi = 0.35$, from~\cite{Dittmann:2014c, Maurath:2015, Bitsch:2014, Hoffmann:2014}.)}  
	\label{fig:apps}
	\end{figure*}

One very promising area where two-fluid suspensions may be used is in ceramic materials. In metal oxide-paraffin oil suspensions used for low-pressure injection molding, it was found that molded green parts that were immersed in water prior to thermal de-binding had an increased strength before they were sintered~\cite{Novak:2000}.  Capillary bridges caused by water immersion or the adsorption of water to these hygroscopic particles can also influence the structure of the sintered parts with complex geometry preventing deformation and rounded edges formed during sintering~\cite{Cetinel:2013}. The increased yield stress can help in the storage of green parts without deformation, but under certain circumstances can also cause flocculation of the metal particles, which introduces defects into the sintered parts~\cite{Novak:2000}.  The flocculation, continued such that a network is formed, can also be utilized to create porous ceramics.  Water, introduced into a paraffin-based suspension, creates a strong sample-spanning network that can maintain its form even as the bulk fluid is removed~\cite{Dittmann:2014c, Dittmann:2013, Dittmann:2012}.  Using this method, homogeneous, crack-free ceramic samples with open porosity above 50\% and an average pore size less than 10~{\textmu}m can be created, as shown in Fig.~\ref{fig:apps}A~\cite{Dittmann:2014c, Dittmann:2013, Dittmann:2012}.
Porous glass samples have also been produced (Fig.~\ref{fig:apps}B), which exhibit a much higher open porosity and more uniform pore size than existing commercial alternatives~\cite{Dittmann:2014, Maurath:2015}.  Most notably, these two samples have nearly the same mechanical strength despite the vastly higher porosity in the capillary suspension formulation~\cite{Maurath:2015}. This manufacturing process is simpler than existing alternatives, results in less waste, can be adapted to different manufacturing techniques, and has the potential to reduce the cost of porous parts.  

Bitsch and coworkers showed that capillary suspensions can be employed to modify water-based pastes used in the manufacturing of electrodes for lithium-ion batteries~\cite{Bitsch:2014}.  Addition of small amounts of octanol increases the low shear viscosity of the electrode slurry without the use of polymeric rheology control agents, which are typically detrimental to the electric properties of the dry electrode~\cite{Bitsch:2014}.  The viscosity at high shear rates is unaffected by the secondary fluid so that the slurries can be processed using well established coating machinery. The particulate network rapidly reforms  after coating. This improves the shape accuracy of the coated thin film and reduces the waste cut-off at the edges (Fig.~\ref{fig:apps}C)~\cite{Bitsch:2014}. The adhesion and electrochemical properties are similar to conventional formulations. The increased shape accuracy may also be advantageous in printed electronics where thin, dense lines can be printed.

Capillary suspensions can also be used to create consumer friendly and low-fat foods.  Water as a secondary fluid can be used in oil-based suspensions to replace additives such as emulsifiers and thickeners, which are viewed unfavorably by consumers, to adjust the mouthfeel of these products as was demonstrated for suspensions with added cocoa solids and starch~\cite{Hoffmann:2014}. The sample-spanning network can also be used in chocolate or other temperature sensitive systems to preserve stability when the continuous phase is molten, as shown in Fig.~\ref{fig:apps}D~\cite{Hoffmann:2014, Killian:2012}. This heat stability is advantageous for the consumer, but can also prevent waste when products, e.g. a chocolate bar, becomes too warm during transport or storage. It may even be possible to replace oil-based suspensions with water-based alternatives to create ultra-low fat foods such as spreads or butter substitutes~\cite{Hoffmann:2014}.

In addition to the previously mentioned applications, there has also been research in other diverse areas. Capillary suspensions have been useful in understanding the formation and structure of hydrate slurries formed from water in mineral oil emulsions~\cite{Zylyftari:2013, Webb:2014}.  These hydrates cause blockages in petroleum pipelines and their prevention is crucial in the drilling of crude oil as these blockages can rapidly increase the pressure in the pipe causing damage to equipment and the environment if the pipe ruptures~\cite{Zylyftari:2013}.  Capillary suspensions have also been used in polymer blends where a percolated particle network can be used to increase the stability of the blends and create stronger composite materials with tunable properties~\cite{Nagarkar:2013, Domenech:2014, Domenech:2015, Wang:2014}.  The use of capillary suspensions was also adapted to create nanocomposite superstructures from molten metal and nanoparticles~\cite{Xu:2013b}. There remain further applications where capillary suspensions may be applied and these should be investigated.

\section{Conclusions and outlook} \label{concl}
Capillary suspensions are ternary particle-liquid-liquid systems where one of the liquid components is a minority phase -- usually present as less than 10\% of the total volume.  The addition of the secondary fluid can transition a suspension from fluid-like or weakly gelled into a strong gel where the viscosity and yield stress increase by several orders of magnitude. These capillary suspensions are named as such due to the strong influence of the capillary force on the network structure and rheological properties. The particles are networked together either as individual particles connected through capillary bridges or as clusters shielding the secondary fluid.  These two states are differentiated according to the wetting behavior of the minority phase: $\theta<90^\circ$ is the pendular state and $\theta>90^\circ$ is the capillary state, where the contact angle $\theta$ is measured using the secondary fluid in an environment provided by the bulk fluid.  Strong capillary state suspensions can not be created using particles with very high contact angles, and this limitation should be explored further.  The influence of the contact angle on the admixture strength in both the capillary and pendular states should also be investigated.

Capillary suspensions are usually formed by adding a small amount of a secondary fluid to an existing suspension and rapidly mixing this ternary system to induce the formation of secondary fluid droplets.  The formation of capillary bridges or clusters of particles requires that these secondary fluid droplets are approximately the same size or smaller than the particles.  While this is often accomplished through rapid mixing and a high specific mechanical energy input upon adding the secondary fluid to a suspension, adding the particles to an emulsion can create a stronger and more uniform capillary suspension.  A uniform capillary network can also be formed from secondary fluid adsorbed to the particle surface. It was observed that each combination of particles and liquids creates a stronger admixture in either the pendular or the capillary state, but not both.  However, given sufficient mixing conditions, it may be possible to create stable suspensions in the non-preferential state.  If this is not possible, we must put forth a hypothesis to explain this limitation.

The properties of the network and microstructure must also be explored further.  While it was proposed that the pendular state is controlled by binary interactions and the capillary state by interactions between tetrahedral or octahedral clusters, secondary fluid droplet polydispersity complicates this simple model.  Direct observations of the network microstructure must be undertaken, particularly in the capillary state.  Additionally, the structure of the sample-spanning network must be directly observed.  The mechanism of network formation and changes during aging must also be investigated.  Finally, the influence of material properties such as the contact angle or surface roughness on the network must be explored. The influence of these material properties on the bulk rheological properties must also be investigated.  Factors such as particle shape can induce network formation at lower solid fractions and may change the strength of the network due to a preferential orientation relative to the secondary fluid bridges.  

Finally, there are many possible applications for these capillary suspensions.  Some of these applications have been partially explored, but this research must be furthered to improve the properties of the final material.  For example, the research with porous materials should be continued to develop technically relevant processing techniques such as injection molding in order to facilitate profitable production of such porous membranes with unprecedented porosity and pore size specifications.  There are many other possible applications, the boundaries of which remain to be explored. Printed electronics is a promising field since the secondary phase can easily replace organic dispersing and rheology control agents, thus providing good processing properties without disturbing electrical end use features. These new applications are only limited by the creativity of the reader.



\section*{Acknowledgments}
The author would like to thank Norbert Willenbacher for fruitful discussions and careful reading of the manuscript.
Financial support was provided from the European Research Council under the European Union's Seventh Framework Program (FP/2007-2013)/ERC Grant Agreement no. 335380.

\makeatletter
\@addtoreset{footnote}{page}
\renewcommand{\thefootnote}{\ifcase\value{footnote}\or*\or**\or($\infty$)\fi}
\newcommand\footnoteref[1]{\protected@xdef\@thefnmark{\ref{#1}}\@footnotemark}
\makeatother

\section*{References and recommended reading\footnote{of special interest\label{n1}}$^,$ \footnote{of outstanding interest\label{n2}}}

\begin{thebibliography}{10}

\expandafter\ifx\csname url\endcsname\relax
  \def\url#1{\texttt{#1}}\fi
\expandafter\ifx\csname urlprefix\endcsname\relax\def\urlprefix{URL }\fi
\expandafter\ifx\csname href\endcsname\relax
  \def\href#1#2{#2} \def\path#1{#1}\fi

\bibitem{Aveyard:2003}
R.~Aveyard, B.P.~Binks, J.H.~Clint, Emulsions stabilised solely by colloidal
  particles, Adv. Colloid Interface Sci. 100--102 (2003) 503--546.

\bibitem{Cattermole:1904}
A.E.~Cattermole, Classification of the metallic constituents of ores, US
  Patent 763259 (1904).

\bibitem{Leonard:1991}
W.G.~Leonard, R.T.~Greer, R.M.~Markuszewski, T.D.~Wheelock, Coal
  desulfurization and deashing by oil agglomeration, Sep. Sci. Technol. 16~(10)
  (1981) 1589 -- 1609.

\bibitem{Stratford:2005}
K.~Stratford, R.~Adhikari, I.~Pagonabarraga, J.C.~Desplat, M.E.~Cates,
  Colloidal jamming at interfaces: {A} route to fluid-bicontinuous gels,
  Science 309~(5744) (2005) 2198--2201.

\bibitem[5$^{\ast\ast}$]{Koos:2011} 
E.~Koos, N.~Willenbacher, Capillary forces in suspension rheology, Science
  331~(6019) (2011) 897--900. [Demonstrates capillary suspensions using a variety of materials and outlines the distinctions between the pendular and capillary states].

\bibitem{Koos:2012b}
E.~Koos, J.~Johannsmeier, L.~Schwebler, N.~Willenbacher, Tuning suspension
  rheology using capillary forces, Soft Matter 8~(24) (2012) 6620--6628.

\bibitem{Dittmann:2013}
J.~Dittmann, E.~Koos, N.~Willenbacher, Ceramic capillary suspensions: {N}ovel
  processing route for macroporous ceramic materials, J. Am. Ceram. Soc. 96~(2)
  (2013) 391--397.

\bibitem[8$^{\ast}$]{Hoffmann:2014} 
S.~Hoffmann, E.~Koos, N.~Willenbacher, Using capillary bridges to tune
  stability and flow behavior of food suspensions, Food Hydrocolloids 40 (2014)
  44--52. [Highlights capillary suspensions in food with a discussion of agglomeration due to adsorption of water on oil-based suspensions of hygroscopic particles and possible applications].

\bibitem{Bloomquist:1940}
C.R.~Blomquist, R.S.~Shutt, Fine particle suspensions in organic liquids,
  Ind. Eng. Chem. 32~(6) (1940) 827--831.

\bibitem{Kao:1975b}
S.V.~Kao, L.E.~Nielsen, C.T.~Hill, Rheology of concentrated suspensions of
  spheres. {II}. {S}uspensions agglomerated by an immiscible second liquid, J.
  Colloid Interface Sci. 53~(3) (1975) 367--373.

\bibitem{Strauch:2012}
S.~Strauch, S.~Herminghaus, Wet granular matter: a truly complex fluid, Soft
  Matter 8~(32) (2012) 8271--8280.

\bibitem{Howe:1955}
P.G.~Howe, D.P.~Benton, I.E.~Puddington, The nature of the interaction
  forces between particles in suspensions of glass spheres in organic liquid
  media, Can. J. Chem. 33~(7) (1955) 1189--1196.

\bibitem{McCulfor:2010}
J.~Mc{C}ulfor, P.~Himes, M.R.~Anklam, The effects of capillary forces on the
  flow properties of glass particle suspensions in mineral oil, AlChE J. 57~(9)
  (2011) 2334--2340.

\bibitem{Cavalier:2002}
K.~Cavalier, F.~Larch\'{e}, Effects of water on the rheological properties of
  calcite suspensions in dioctylphthalate, Colloids Surf., A 197~(1-3) (2002)
  173--181.

\bibitem{Butt:2009}
H.J.~Butt, M.~Kappl, Normal capillary forces, Adv. Colloid Interface Sci.
  146~(1--2) (2009) 48--60.

\bibitem{Reis:2010}
P.M.~Reis, S.~Jung, J.M.~Aristoff, R.~Stocker, How cats lap: {W}ater uptake
  by \emph{Felis catus}, Science 330~(6008) (2010) 1231--1234.

\bibitem{Duan:2010}
H.~Duan, K.K.~Berggren, Directed self-assembly at the 10 nm scale by using
  capillary force-induced nanocohesion, NanoLett 10~(9) (2010) 3710--3716.

\bibitem{Herminghaus:2005}
S.~Herminghaus, Dynamics of wet granular matter, Adv. Phys. 54~(3) (2005)
  221--244.

\bibitem{Willett:2000}
C.D.~Willett, M.J.~Adams, S.A.~Johnson, J.P.K.~Seville, Capillary bridges
  between two spherical bodies, Langmuir 16~(24) (2000) 9396--9405.

\bibitem{Butt:2008}
H.J.~Butt, Capillary forces: Influence of roughness and heterogeneity,
  Langmuir 24~(9) (2008) 4715--4721.

\bibitem[21$^{\ast}$]{Manoharan:2003} 
V.N.~Manoharan, M.T.~Elsesser, D.J.~Pine, Dense packing and symmetry in
  small clusters of microspheres, Science 301~(5632) (2003) 483--487. [Demonstration of colloidal clusters produced from Pickering emulsions with a description of the cluster geometry].

\bibitem{Lauga:2004}
E.~Lauga, M.P.~Brenner, Evaporation-driven assembly of colloidal particles,
  Phys. Rev. Lett. 93~(23) (2004) 238301.

\bibitem{Cho:2008}
Y.S. Cho, G.R. Yi, S.H. Kim, M.T.~Elsesser, D.R.~Breed, S.M.~Yang,
  Homogeneous and heterogeneous binary colloidal clusters formed by
  evaporation-induced self-assembly inside droplets, J. Colloid Interface Sci.
  318~(1) (2008) 124--133.

\bibitem[24$^{\ast}$]{Koos:2012} 
E.~Koos, N.~Willenbacher, Particle configuration and gelation in capillary
  suspensions, Soft Matter 8~(14) (2012) 3988--3994. [Model of clusters forming the basic building blocks of capillary state suspensions with predictions for the rheological properties].

\bibitem{Peng:2013}
B.~Peng, F.~Smallenburg, A.~Imhof, M.~Dijkstra, A.~{van B}laaderen, Colloidal
  clusters by using emulsions and dumbbell-shaped particles: {E}xperiments and
  simulation, Angew. Chem. Int. Ed. 52~(26) (2013) 6709--6712.

\bibitem{Brugarolas:2013}
T.~Brugarolas, F.~Tu, D.~Lee, Directed assembly of particles using microfluidic
  droplets and bubbles, Soft Matter 9~(38) (2013) 9046--9058.

\bibitem{Yi:2002}
G.R.~Yi, V.N.~Manoharan, S.~Klein, K.R.~Brzezinska, D.J.~Pine, F.F.~Lange,
  S.M.~Yang, Monodisperse micrometer‐scale spherical assemblies of polymer
  particles, AdvMater 14~(16) (2002) 1137--1140.

\bibitem{Wang:2012}
Y.~Wang, Y.~Wang, D.R.~Breed, V.N.~Manoharan, L.~Feng, A.D.~Hollingsworth,
  M.~Weck, D.J.~Pine, Colloids with valence and specific directional bonding,
  Nature 491~(7422) (2012) 51--55.

\bibitem{Wagner:2008}
C.S.~Wagner, Y.~Lu, A.~Wittemann, Preparation of submicrometer-sized clusters
  from polymer spheres using ultrasonication, Langmuir 24~(21) (2008)
  12126--12128.

\bibitem{Wagner:2010}
C.S.~Wagner, B.~Fischer, M.~May, A.~Wittemann, Templated assembly of polymer
  particles into mesoscopic clusters with well-defined configurations, Colloid.
  Polym. Sci. 288~(5) (2010) 487--498.

\bibitem{Wagner:2012}
C.S.~Wagner, A.~Fortini, E.~Hofmann, T.~Lunkenbein, M.~Schmidt, A.~Wittemann,
  Particle nanosomes with tailored silhouettes, Soft Matter 8~(6) (2012)
  1928--1933.

\bibitem[32$^{\ast}$]{Schwarz:2011} 
I.~Schwarz, A.~Fortini, C.S.~Wagner, A.~Wittemann, M.~Schmidt, {M}onte {C}arlo
  computer simulations and electron microscopy of colloidal cluster formation
  via emulsion droplet evaporation, J. Chem. Phys. 135~(24) (2011) 244501. [Experiments and simulations of colloidal clusters with attractive interactions having non-minimal second moment configurations].  

\bibitem{Arkus:2009}
N.~Arkus, V.N.~Manoharan, M.P.~Brenner, Minimal energy clusters of hard
  spheres with short range attractions, Phys. Rev. Lett. 103~(11) (2009)
  118303.

\bibitem{Meng:2010}
G.~Meng, N.~Arkus, M.P.~Brenner, V.N.~Manoharan, The free-energy landscape of
  clusters of attractive hard spheres, Science 327~(5965) (2010) 560--563.

\bibitem{Cho:2005}
Y.S.~Cho, G.R.~Yi, J.M.~Lim, S.H.~Kim, V.N.~Manoharan, D.J.~Pine, S.M.~Yang, 
  Self-organization of bidisperse colloids in water droplets, J. Am.
  Chem. Soc. 127~(45) (2005) 15968---15975.

\bibitem{Kim:2008}
S.H.~Kim, G.R.~Yi, K.H.~Kim, S.M.~Yang, Photocurable {P}ickering emulsion
  for colloidal particles with structural complexity, Langmuir 24~(6) (2008)
  2365--2371.

\bibitem{Kraft:2012}
D.J.~Kraft, R.~Ni, F.~Smallenburg, M.~Hermes, K.~Yoon, D.A.~Weitz, A.~van
  {B}laaderen, J.~Groenewold, M.~Dijkstra, W.K.~Kegel, Surface roughness
  directed self-assembly of patchy particles into colloidal micelles,
  ProcNatlAcadSciUSA 109~(27) (2012) 10787--10792.

\bibitem{Mao:2012}
X.~Mao, Q.~Chen, S.~Granick, Entropy favours open colloidal lattices, NatMater
  12 (2012) 217--222.

\bibitem{Forster:2011}
J.D.~Forster, J.G.~Park, M.~Mittal, H.~Noh, C.F.~Schreck, C.S.~O'{H}ern,
  H.~Cao, E.M.~Furst, E.R.~Dufresne, Assembly of optical-scale dumbbells into
  dense photonic crystals, ACS Nano 5~(8) (2011) 6695--6700.

\bibitem{Wang:2014b}
Y.~Wang, Y.~Wang, X.~Zheng, G.R.~Yi, S.~Sacanna, D.J.~Pine, M.~Weck,
  Three-dimensional lock and key colloids, JAmChemSoc 136~(19) (2014)
  6866--6869.

\bibitem{Brakke:1992}
K.A.~Brakke, The surface evolver, Exp. Math 1~(2) (1992) 141--165.

\bibitem{Hijnen:2014}
N.~Hijnen, P.S.~Clegg, Colloidal aggregation in mixtures of partially miscible
  liquids by shear-induced capillary bridges, Langmuir 30~(20) (2014)
  5763--5770.

\bibitem{Heidlebaugh:2014}
S.J.~Heidlebaugh, T.~Domenech, S.V.~Iasella, S.S.~Velankar, Aggregation and
  separation in ternary particle/oil/water systems with fully-wettable
  particles, Langmuir 30~(1) (2014) 63--74.

\bibitem[44$^{\ast}$]{Domenech:2014} 
T.~Domenech, S.S.~Velankar, Capillary driven percolating networks in ternary
  blends of immiscible polymers and silica particles, Rheol. Acta DOI:
  10.1007/s00397-014-0776-0. [Examination of pendular state suspensions produced using various mixing conditions demonstrating the importance of small, uniform droplets on material strength].

\bibitem[45$^{\ast\ast}$]{Domenech:2015}  
T.~Domenech, S.S.~Velankar, On the rheology of pendular gels and morphological
  developments in paste-like ternary systems based on capillary attraction, Unpublished manuscript. [Investigation of the rheological differences of pendular state suspensions with increasing secondary fluid and particle volume fractions showing changes with the transition from bridges to cluster].
  
\bibitem{Kaur:2010}
S.~Kaur, L.G.~Leal, Drop deformation and break-up in concentrated suspensions,
  J. Rheol. 54~(5) (2010) 981--1008.

\bibitem{Koos:2014}
E.~Koos, W.~Kannowade, N.~Willenbacher, Restructuring and aging in a capillary
  suspension, Rheol. Acta DOI: 10.1007/s00397-014-0805-z.

\bibitem[48$^{\ast\ast}$]{Fortini:2012} 
A.~Fortini, Clustering and gelation of hard spheres induced by the {P}ickering
effect, Phys. Rev. E 85~(4) (2012) 040401. [Brownian dynamics simulation of particles agglomerated through a non-preferentially wetting fluid with estimations of the transient scattering functions and dimensionality].

\bibitem{Gogelein:2010}
C.~G\"{o}gelein, M.~Brinkmann, M.~Schr\"{o}ter, S.~Herminghaus, Controlling the
  formation of capillary bridges in binary liquid mixtures, Langmuir 26~(22)
  (2010) 17184--17189.

\bibitem{Cheng:2012}
T.L.~Cheng, Y.U.~Wang, Spontaneous formation of stable capillary bridges for
  firming compact colloidal microstructures in phase separating liquids: {A}
  computational study, Langmuir 28~(5) (2012) 2696--2703.

\bibitem{Novak:2000}
S.~Novak, A.~Dakskobler, V.~Ribitsch, The effect of water on the behaviour of
  alumina--paraffin suspensions for low-pressure injection moulding ({LPIM}),
  J. Eur. Ceram. Soc. 20~(12) (2000) 2175--2181.

\bibitem{Cetinel:2013}
F.A.~\c{C}etinel, W.~Bauer, Ceramic micro parts. {P}art {2}: {P}rocess-related
  factors influencing surface finish and shape retention during thermal
  debinding, J. Eur. Ceram. Soc. 33~(15-16) (2013) 3135--3144.
  
\bibitem{Dittmann:2014c}
J.~Dittmann, N.~Willenbacher, Tailoring the mechanical strength of 
highly porous sintering materials with capillary suspensions, Unpublished manuscript.

  \bibitem{Dittmann:2012}
J.~Dittmann, E.~Koos, N.~Willenbacher, B.~Hochstein, Process for producing a
  porous ceramic and porous ceramic obtainable thereby, International Patent WO
  2013004336 A1 (2012).

\bibitem{Maurath:2015}
J.~Maurath, J.~Dittmann, N.~Schultz, N.~Willenbacher, Fabrication of highly porous 
glass membranes by using capillary suspension processing, Unpublished manuscript.

\bibitem{Dittmann:2014}
J.~Dittmann, E.~Koos, N.~Willenbacher, B.~Hochstein, Method for producing
  porous glass bodies, International Patent WO 2014056616 A1 (2014).

\bibitem{Bitsch:2014}
B.~Bitsch, J.~Dittmann, M.~Schmitt, P.~Scharfer, W.~Schabel, N.~Willenbacher, A
  novel slurry concept for the fabrication of lithium-ion battery electrodes
  with beneficial properties, J. Power Sources 265 (2014) 81--90.

\bibitem{Killian:2012}
L.A.~Killian, J.N.~Coupland, Manufacture and application of water-in-oil
  emulsions to induce the aggregation of sucrose crystals in oil: {A} model for
  melt-resistant chocolate, Food Biophys. 7~(2) (2012) 124--131.

\bibitem{Zylyftari:2013}
G.~Zylyftari, J.W.~Lee, J.F.~Morris, Salt effects on thermodynamic and
  rheological properties of hydrate forming emulsions, Chem. Eng. Sci. 95
  (2013) 148--160.

\bibitem{Webb:2014}
E.B.~Webb, C.A.~Koh, M.W.~Liberatore, High pressure rheology of hydrate
  slurries formed from water-in- mineral oil emulsions, Ind. Eng. Chem. Res.
  53~(17) (2014) 6998---7007.

\bibitem{Nagarkar:2013}
S.P.~Nagarkar, S.S.~Velankar, Rheology and morphology of model immiscible
  polymer blends with monodisperse spherical particles at the interface, J.
  Rheol. 57~(3) (2013) 901--925.

\bibitem{Wang:2014}
D.~Wang, X.~Wang, Y.~Yuan, W.~Li, H.~Tian, S.~Zhao, Increasing the apparent
  shear viscosity of polymer composites by uptake of a small amount of water,
  RSC Adv. 4~(47) (2014) 24686--24691.

\bibitem{Xu:2013b}
J.Q.~Xu, L.Y.~Chen, H.~Choi, H.~Konish, X.C.~Li, Assembly of metals and nanoparticles
  into novel nanocomposite superstructures, Sci. Rep. 3 (2013) 1730.

\end{thebibliography}

%
%

\end{document}